\documentclass[authoryear,preprint]{elsarticle}

\usepackage[T1]{fontenc}
\usepackage[utf8]{inputenc}
\usepackage{lmodern}

\usepackage{natbib} 
\bibpunct{[}{]}{;}{n}{,}{,} 

\usepackage{graphicx}
\usepackage{color} 
\usepackage{setspace}

\usepackage{tabularx}
\usepackage{multirow} 

\usepackage{graphicx}  
\usepackage{natbib} 
\usepackage{xspace} 
\usepackage{color}

\usepackage{amssymb}
\usepackage{amsmath}
\usepackage{caption}

\newcommand{\Cplusplus}{{\rmfamily C\raise.22ex\hbox{\small ++} }}

\usepackage{hyperref}
\makeatletter
\providecommand{\doi}[1]{%
  \begingroup
    \let\bibinfo\@secondoftwo
    \urlstyle{rm}%
    \href{http://dx.doi.org/#1}{%
      doi:\discretionary{}{}{}%
      \nolinkurl{#1}%
    }%
  \endgroup
}
\makeatother

\sloppypar

\graphicspath{{.}{figs/}}

\journal{Journal of Parallel and Distributed Computing}

\begin{document}
\begin{frontmatter}
\title {A Survey of High Level Frameworks in Block-Structured Adaptive Mesh Refinement Packages}
\author [1]{Anshu Dubey\corref{cor}}
\ead{adubey@lbl.gov}
\author [1]{Ann Almgren}
\author [1]{John Bell}
\author[8]{Martin Berzins}
\author [4,5]{Steve Brandt}
\author[9]{Greg Bryan}
\author[1,10]{Phillip Colella}
\author[1]{Daniel Graves}
\author [1]{Michael Lijewski}
\author [4]{Frank Löffler}
\author [3]{Brian O'Shea}
\author [6,7,4]{Erik Schnetter}
\author [1]{Brian Van Straalen}
\author [2]{Klaus Weide}

\cortext[cor]{Corresponding author\\~~~~~ One Cyclotron
  Road, Mailstop 50A1148\\~~~~Berkeley, CA 94720\\~~~~~Phone: 510-486-5242 \\ ~~~~~~Fax: 510-486-6900}

\address[1]{Computational Research Division, \\Lawrence Berkeley
  National Laboratory,\\ Berkeley, CA 94720} 
\address[2]{Flash Center for Computational Science, \\The University of
  Chicago, \\Chicago 60637}
\address[3]{Department of Physics and Astronomy, Lyman Briggs College,\\
  and the Institute for Cyber-Enabled Research,\\ Michigan State
  University, \\East Lansing, MI 48824, USA}
\address[4]{Center for Computation and Technology, \\Louisiana State University, \\Baton Rouge, LA 70803, USA}
\address[5]{Department of Computer Science, \\Louisiana State University, \\Baton Rouge, LA 70803, USA}
\address[6]{Perimeter Institute for Theoretical Physics, \\Waterloo, ON N2L 2Y5, Canada}
\address[7]{Department of Physics, University of Guelph,\\ Guelph, ON N1G 2W1, Canada}
\address[8]{Mathematics and School of Computing, University of Utah, \\Salt Lake City, UT 84112}
\address[9]{Department of Astronomy, Columbia University,\\ New York, NY, 10025, USA}
\address[10]{Computer Science Department, University of
  California,\\Berkeley, CA 94720}

\date{April 18, 2014}

\begin{abstract}
Over the last decade block-structured adaptive mesh refinement (SAMR) has
found increasing use in large, publicly available codes and
frameworks. SAMR frameworks have evolved along different paths. Some
have stayed focused on specific domain areas, others have pursued a
more general functionality, providing the building blocks for a larger
variety of applications. In this survey paper we examine a
representative set of SAMR packages and SAMR-based codes that have been
in existence for half a decade or more, have a reasonably sized and
active user base outside of their home institutions, and are publicly
available. The set consists of a mix of SAMR packages and application
codes that cover a broad range of scientific domains. We look at their
high-level frameworks, and their approach to dealing with the advent
of radical changes in hardware architecture.  The codes included in
this survey are BoxLib, Cactus, Chombo, Enzo, FLASH, and Uintah.
\end{abstract}


\end{frontmatter}  
\newpage
\section{Introduction\label{sec:intro}}

Block-structured adaptive mesh refinement (SAMR) \citep{Berger1984,Berger1989} 
first appeared as a computational technique almost 30 years ago; 
since then it has been used in many individual research codes and, 
increasingly over the last decade, in large, publicly available code
frameworks and application codes. The first uses of SAMR focused
almost entirely on  explicit methods for compressible 
hydrodynamics, and these types of problems motivated the building of
many of the large code frameworks.   SAMR frameworks have evolved along
different paths. Some have stayed focused on specific domain areas,
adding large amounts of functionality and problem-specific physics
modules that are relevant to those applications.  Examples of these
include AstroBEAR \citep{AstroBEAR}, CRASH \citep {Holst2011},
Cactus\citep{Goodale:2002a,Cactuscode:web},  Enzo
\citep{Enzo2014,Enzocode:web}, FLASH \citep{Fryxell2000,Dubey2009},   
Overture \citep{Overture:web}, PLUTO \citep{Mignone2012}, \citep
{RAMSES2002}, and Uintah \citep{uintah2,csafe3}. Other frameworks have
pursued a more general functionality, providing the building blocks
for a larger variety of applications while enabling domain-specific
codes to be built using that framework. As an example, while almost every
SAMR framework can be used to solve systems of hyperbolic
conservation laws explicitly, not all frameworks include the functionality to
solve elliptic equations accurately on the entire hierarchy or a
subset of levels. Examples of frameworks constructed specifically for solving
hyperbolic conservation laws include AMROC \citep{deiterding2002amroc} 
and AMRClaw \citep{AMRCLAW},  both based on the wave propagation algorithms of R. LeVeque.
Extensions of AMRClaw include GeoClaw \citep{GeoClaw}, the widely used tsunami simulation tool.
BoxLib \citep{Boxlib}, Chombo \citep{Chombo2009}, Jasmine \citep{jasmin2010}
and SAMRAI \citep{samrai2008,hornung2002} are more general in that
they supply full functionality for solving equation sets  containing
hyperbolic, parabolic and elliptic equations, and facilitate the
development of codes for simulating a wide variety of different
applications. PARAMESH \citep{MacNeice2000} supplies only the mesh
management capability and  as such is equation-independent.  A more
comprehensive list of codes that use SAMR, and other useful adaptive
mesh refinement (AMR) resources can be found at \citep{AMRpage}.  

SAMR codes all rely on the same fundamental concept, viz. that
the solution can be computed in different regions of the domain with 
different spatial resolutions, where each region at a particular resolution
has a logically rectangular structure.   In some SAMR codes the data is
organized by level, so that the description of the hierarchy is fundamentally 
defined by the union of blocks at each level; while others organize
their data with unique parent-child relationships.   Along with the
spatial decomposition, different codes solving time-dependent
equations make different assumptions about the time stepping, i.e.,
whether grids at all levels advance at the same time step, or grids
advance with a time step unique to their level.  Finally, even when
frameworks are used to solve exactly the same equations with  exactly
the same algorithm, the performance can vary due to the fact that
different frameworks are written in different languages, with
different choices of data layout, implementation details,
etc. However, despite their differences in infrastructure and target
domain applications, the codes have many aspects that are similar, and
many of these codes follow a set of very similar software engineering
practices.   

In this survey paper we examine a representative set of SAMR packages
and SAMR-based codes that: (1) have been in existence for half a decade
or more, (2) have a reasonably sized and active user base outside of
their home institutions, and most importantly, (3) are publicly
available to any interested user. In selecting the codes we have taken
care to include variations in spatial and temporal refinement
practices, load distribution and meta-information
management. Therefore, we have octree and patch based SAMR, no subcycling and
subcycling done in different ways, load distribution on a level by
level or all levels at once,  and globally replicated meta-data or
local view. Additionally, the set covers a broad range of scientific
domains that use SAMR technology in different ways. We look at their
high-level frameworks, consider the trade-offs between various
approaches, and the challenges posed by the advent of radical
changes in hardware architecture.  The codes studied 
in detail in this survey are BoxLib, Cactus, Chombo, Enzo, FLASH, and
Uintah. The application domains covered by a union of these codes
include astrophysics, cosmology, general relativity, combustion,
climate science, subsurface flow, turbulence, fluid-structure
interactions, plasma physics, and particle accelerators.

\section{Overview of the Codes\label{sec:overview}}
BoxLib is primarily a framework for building massively parallel
SAMR applications. The goals of the BoxLib framework are 
twofold: first, to support the rapid development, implementation and
testing of new algorithms in a massively  parallel SAMR framework; and
second, to provide the basis for large-scale domain-specific
simulation codes to be used for numerical investigations of phenomena
in fields such as astrophysics, cosmology, subsurface flow, turbulent
combustion, and any other field which can be fundamentally described
by time-dependent PDE's (with additional source terms, constraints,
etc).   In accordance with both goals, the core remains relatively
agile and agnostic, and is not tied to a particular time-stepping or
spatial discretization strategy, or to a particular set of physics
packages. 

That said, while BoxLib itself supplies a very general
capability for solving time-dependent PDE's on an adaptive mesh
hierarchy, there are a number of large, domain-specific BoxLib-based
application codes in scientific use today.  The most widely used
include  CASTRO \citep{CASTRO,CASTRO2,CASTRO3} for fully compressible
radiation-hydrodynamics;  MAESTRO \citep{MAESTRO}, for low Mach number
astrophysical flows;  Nyx \citep{Nyx} for cosmological applications;
LMC \cite{LMC} for low Mach number combustion; and the structured
grid component of the Amanzi code for modeling subsurface flow
\citep{PMAMR}.  

Chombo is an offshoot of the BoxLib framework, having branched off from BoxLib in 1998.  
As a result Chombo shares many features with BoxLib, including 
the hybrid \Cplusplus / Fortran approach, and the separation of
concerns between the parts that are best handled in  \Cplusplus
(abstractions, memory management, I/O, flow control) and  Fortran
dialects (loop parallelism, stencil computations). Chombo articulates
its APIs explicitly for easier plugin by the client application
code. It has also diverged from BoxLib in the design of its data containers.
Chombo currently supports applications in a range of disciplines, including the following:  
MHD for tokamaks using all-speed projection methods, and  large eddy
simulations of wind turbines at KAUST; compressible CFD +
collision-less particle cosmology simulations ({ CHARM} code,  
\cite{miniatiColella:2007,miniatiMartin});
 physics of the solar wind and its interaction with the interstellar
 medium using compressible hyperbolic CFD + electromagnetic and
 kinetic effects ({ MS-FLUKSS} code
 \cite{2012Sci...336.1291M,2013ApJ...772....2P,2012ApJ...750L...4P});
 general astrophysics modeling ({ PLUTO} code
\cite{pluto});
 astrophysical MHD turbulence ({ ORION} code 
\cite{orion});
 SF Bay and Delta hydrology modeling – shallow water ({ Realm} code 
\cite{realm});  plasma-wakefield accelerators -- compressible viscous flow at LBNL;
 blood flow in cerebral arteries -- fluid / solid coupling (UCB ParLab project \cite{deschampsETAL:2004});
 pore-scale subsurface reacting flow ({ Chombo-Crunch} code, 
\cite{trebotich:2011}); 
 conjugate heat transfer in nuclear reactors 
\cite{crockettETAL:2010}); 
 4D gyrokinetic models of tokamak edge plasmas ({ COGENT} code, 
\cite{dorrETALSciDAC:2010,colellaETALMapped:2011});
 land ice model for climate simulation  ({ BISICLES} code, 
\cite{martinETAL:2011}); and atmospheric models for climate simulation – low-Mach number CFD 
at U. Michigan.

Cactus~\cite{Goodale:2002a,Cactuscode:web} was designed
as a general-purpose software framework for high-performance
computing with AMR as one of its features. The first set of applications 
that used the framework were astrophysical simulations of compact
objects involving general relativity (GR) such as black holes and
neutron stars. These scenarios require high resolution inside and near
the compact objects, and at the same time need to track gravitational
waves propagating to large distances of several hundred times the
radii of the compact objects. This leads very naturally to the use of
AMR. In GR, gravitational effects are described via hyperbolic
(wave-type) equations, propagating at the speed of light. This removes
the need for an elliptic solver that is necessary in Newtonian gravity
to calculate the gravitational potential. While the Cactus framework
is generic, its most prominent user today is the Einstein
Toolkit~\cite{Loffler:2011ay,Zilhao:2013aa,EinsteinToolkit:web}, a
large set of physics modules for relativistic astrophysics
simulations.  The Einstein Toolkit includes  modules for solving the
Einstein equations and relativistic magneto-hydrodynamics, as well as
modules for initial conditions, analysis, and so on.  

Enzo is a standalone application code~\citep{Enzo2014,Enzocode:web} that was originally designed to
simulate the formation of large-scale cosmological structure, such as
clusters of galaxies and the intergalactic medium.  Since the
formation of structures in the Universe is a process that is driven by
gravitational collapse, the study of this phenomenon naturally
requires high resolution in both space and time, making AMR
a logical choice.  Since the first version of Enzo was written in
1996, the user base has grown to include roughly 100 researchers studying a
variety of astrophysical phenomena, including galaxies, galaxy
clusters, the interstellar and intergalactic media, turbulence, and star
formation in the early universe and in our own Galaxy.  Because of
this growth, a wide range of capabilities have been added to the Enzo
code, including a range of hydrodynamic and magnetohydrodynamic
solvers, implicit flux-limited radiation diffusion and explicit
radiation transport with a ray-casting method, optically-thin and
thick radiative cooling, prescriptions for active particles and
passive tracer particles, and a wide variety of problem types that users can expand
upon to pursue their own interests. 

FLASH \citep{Dubey2009, Dubey2008, Fryxell2000} was originally designed
for simulating astrophysical phenomena dominated by compressible
reactive flows. The target applications also had multiple physical scales and
therefore required AMR, which was provided by the octree-based
PARAMESH package \citep{MacNeice2000}. Though PARAMESH supports
subcycling in time FLASH does not.  FLASH underwent three iterations of
infrastructure refactoring, and the resultant architecture has enabled the
code to be readily extensible. As a result, capabilities have been added
to the code to make it useful for such disparate communities as
cosmology, astrophysics,  high-energy-density physics, computational
fluid dynamics, and fluid-structure interactions. FLASH's capabilities
include solvers for hydrodynamics, magnetohydrodynamics, self-gravity,
radiation in the flux-limited diffusion formulation, several specialized
equations of state, material properties such as magnetic resistivity
and conductivity, several source terms including nuclear burning and
laser driver, tracer and active particles, and immersed boundaries.

The Uintah software was initially a direct outcome from 
the University of Utah DOE Center for the Simulation of Accidental Fires
and Explosions (C-SAFE)~\cite{csafe3} 
that  focused on providing state-of-the-art, science-based tools
for the numerical simulation of accidental fires and explosions.
The Uintah framework allows chemistry and engineering physics to be
fully coupled with nonlinear solvers and  visualization packages.  
The Uintah open-source (MIT License) software has been widely ported and used for many different types of 
problems involving fluid, solid, and fluid-structure interaction problems.
The present status of Uintah is described by \cite{MBUintahreport}.

Uintah presently contains four main simulation algorithms, or components: 
1) the ICE~\cite{kashiwa2003,kashiwaetal96} compressible multi-material
finite-volume CFD component, 2) the particle-based Material Point Method
(MPM)~\cite{sulskycmame} for structural mechanics, 3) the combined
fluid-structure interaction (FSI) algorithm MPM-ICE~\cite{spain1,spain2,fourthmit}, and 4)
the Arches turbulent reacting CFD component \cite{Arches,Smith}
that was designed for simulation of turbulent reacting flows
with participating media radiation.  Arches is a three-dimensional, Large Eddy
Simulation (LES) code that uses a low-Mach number
variable density formulation to simulate heat, mass, and
momentum transport in reacting flows.
Uintah has been used, and is being used, for many different
projects such as angiogenesis, tissue
engineering, heart injury modeling, blast-wave simulation,
semiconductor design, and multi-scale materials research)
\cite{MBUintahreport}.

\section {Frameworks \label{sec:frameworks}}
There are a number of similarities between the six codes / software frameworks
(which from now on we will call ``codes'') described in this paper.
Each of these codes provides some generic support for SAMR
applications as well as more specialized support for specific
applications. Since the codes detailed in the survey come from different
disciplines, groups, and scientific domains they each use various
terms in their own different ways. In order to facilitate the
discussion we override individual code's usage and adhere to the 
following terminology:
\begin{itemize}
\item {\bf cell}: the smallest unit of discretized domain
\item{\bf mesh/grid}: generic way of describing the discretized domain
\item {\bf block}: logically rectangular collection of cells
\item {\bf patch}: a collection of contiguous cells of the same size
  (at the same refinement level),  patches may be subdivided into blocks
\item{\bf active cells}: cells in a block that are updated
  when an operator is applied to the block
\item{\bf guard cells}: halo of cells surrounding the active cells
  that are needed for computation of an operator, but are not updated
  by the operator. 
\item{\bf level}: union of blocks that have the same cell size
\item{\bf framework}: the infrastructure backbone of the code
\item{\bf component}: an encapsulated stand-alone functionality within
  the code
\end{itemize}

All codes perform domain decomposition into blocks. 
BoxLib, Cactus, Chombo and Uintah are perhaps the most general frameworks, in
that the bulk of the software capability is not tied to a particular
application.  Enzo is perhaps the most specific in that it is specifically
designed for astrophysical and cosmological applications. As such, the Enzo
release contains numerous modules for specific processes such as star formation
and feedback. FLASH lies between BoxLib/Cactus/Chombo/Uintah and Enzo;
it has extensive physics-independent infrastructure, but also
includes physics-specific modules and solvers for a variety of applications in
its releases.  All of the codes support finite difference / finite volume methods 
in 1, 2 or 3 dimensions in Cartesian coordinates and all codes except Cactus support particle
and particle/mesh algorithms, with both active and passive
particles. Support is provided for data that lives on cell centers,
faces, edges, or nodes. Except FLASH, all other codes support
subcycling in time. The original parallelization model in these codes,
as with most codes of similar vintage was distributed memory with MPI,
though now they have various degrees of hybrid parallelization as listed
in Table 
\ref{tab:features}. 

In all of the codes, explicit hyperbolic solvers act upon individual
blocks with no knowledge of other blocks once the guard cell data have
been filled from blocks at the same or coarser levels as appropriate.
Explicit refluxing occurs at coarse-fine boundaries to correctly
update the solution The implicit and semi-implicit solvers place more demands on
the communications and different codes handle them differently. An
interesting observation is that the original implicit and semi-implicit
solvers came in the form of geometric multigrid in those codes
that had any. As the codes started supporting capabilities more
demanding of such solvers they started to provide interfaces to the
readily available capabilities from libraries such as PETSc and
Hypre. Note that because of AMR, and therefore the presence of
fine-coarse boundaries,  these interfaces are non-trivial and involve
considerable effort to do correctly. 
 Chombo, Flash and Uintah have support for fluid/structure interaction
 -- in Chombo, embedded boundaries are used to represent the
 fluid/solid interface; FLASH uses an immersed boundary
 representation, and Uintah uses the particle-based Material Point
 Method (MPM) for structural modeling.

\subsection{BoxLib \label{sec:boxlib}}
The main branch of BoxLib is a combination of \Cplusplus \slash
Fortran90 software; a newer branch of BoxLib is written in pure
Fortran90 but currently supports only non-subcyling algorithms. 
BoxLib contains extensive support for both explicit and implicit grid-based operations 
as well as particles on hierarchical adaptive meshes.  Single-level and multi-level
multigrid solvers are included for cell-based and node-based data.  
The fundamental parallel abstraction in both the \Cplusplus  and 
Fortran90 versions is the MultiFab, which holds the data on the union
of blocks at a level.  A MultiFab is composed of multiple FABs
(Fortran Array Boxes); each FAB is an array of data on a single block.  
During each MultiFab operation the FABs composing that MultiFab 
are distributed among the nodes; MultiFabs at each level of refinement are distributed
independently.  
Each node holds {\it meta-data} that is needed to fully specify the geometry 
and processor assignments of the MultiFabs.  
The meta-data can be used to dynamically evaluate the necessary
communication patterns for sharing data amongst processors in order
to optimize communications patterns within the algorithm.  

The scaling behavior of BoxLib depends strongly on the algorithm being implemented.
Solving only a system of hyperbolic conservation laws, CASTRO has achieved excellent
weak scaling to $196$K cores on the jaguarpf machine at the Oak Ridge Leadership 
Computing Facility (OLCF) using only MPI-based parallelism. 
Good scaling of linear solves is known to be much more difficult to achieve. 
Recent scaling results of the Nyx code, which relies on multigrid solves of the
Poisson equation for self-gravity, demonstrate excellent scaling up to 49K cores on 
the Hopper machine at NERSC.

\subsection{Chombo \label{sec:chombo}}
 Many features in Chombo, such as hybrid language model, are
similar to BoxLib.  Chombo's union of blocks is called a {\tt
  BoxLayout} (with a specialization being a  {\tt 
  DisjointBoxLayout}). These maintain the mapping of blocks to
compute elements (nodes, or cores). This {\it meta-data} is replicated
across all MPI ranks redundantly.  In cases with extreme block counts
(O($10^6$) blocks) this meta-data is compressed \cite{van2011petascale}. 
Meta-data is shared at the thread-level.  Cell-based
refinment codes have a different parameter space to operate in and can
have significantly higher meta-data costs in return for higher
floating-point efficiency. 

   Chombo keeps the {\tt FAB} (Fortran Array Box) data member from BoxLib, but it is
templated on data type and data centering.  Instead of {\tt MultiFAB},
Chombo has a hierarchy of templated data holders ( {\tt LayoutData},
{\tt BoxLayoutData}, and {\tt LevelData}). An important reason for the
templated data holder design is to provide a common code base that
also supports the {\tt EBCellFAB}, a cell-centered finite volume data
holder that supports embedded boundary algorithms
\cite{apdec:ebchombo,apdec:ebamrtools,apdec:ebamrgodunov,gravesETAL:2006},  
and the {\tt BinFAB} to support Particle-In-Cell algorithms.  
Chombo also supports mapped multiblock domains and mixed-dimension domains up to 6D.

Chombo has built-in diagnostics to track time in functions (serial and
parallel), memory leak tracking, memory tracing, sampling profilers,
and  the ability to trap and attach a native debugger to a running job
as part of the default build environment without requiring third party
packages.  Chombo has demonstrated scaling behavior of both its
hyperbolic gas dynamics codes and its multigrid  elliptic solver to
$196$K cores on the Jaguar machine at the Oak Ridge Leadership
Computing Facility (OLCF) using flat MPI-based parallelism
\cite{van2011petascale}.  

\subsection{Cactus / Carpet \label{sec:cactus}}

Cactus modules consist of routines targeting blocks, where the
core Cactus framework manages when and how these routines
are called. The framework itself does not provide parallelism or AMR;
these are instead implemented via a special \emph{driver} component.
These days, \emph{Carpet}~\cite{Schnetter:2003rb,
Schnetter:2006pg,CarpetCode:web} is the only widely-used driver. In
principle, it would be possible to replace Carpet by an alternative
driver that provides, e.g., a different AMR algorithm; if that new
driver adhered to the existing interfaces, neither the framework nor
existing physics modules would need to be modified.

Carpet supports two ways to define the grid hierarchy. Application
components can explicitly describe locations, shapes, and depths of
refined regions, which is useful, e.g., when tracking black holes or
stars. Alternatively, the application can mark individual cells
for refinement, and Carpet will then employ a tiling method
(implemented in parallel) to find an efficient grid structure that
encloses all marked points. 


Cactus relies on a domain-specific language (DSL) describing its
modules~\cite{Seidel:2010bb}. This DSL describes the 
schedule (workflow) of tasks in a Cactus computation, the grid
functions (a distributed data structure which contains the values of a
field on every point of the grid), and parameter files. The framework
provides an API that lets infrastructure components (e.g., the driver)
query this information. One distinguishing feature of Cactus is that
the components self-assemble -- the information provided via this
DSL is rich enough that components can simply be added to an existing
simulation without explicitly specifying inter-component data flow.
(Note that components have to be designed with this in mind.)
This framework design requires Cactus applications to modularize
functionality to a high degree. 

Cactus is designed such that the executable is almost always compiled by the user,
from the source code of Cactus and its modules, and, optionally, additional
private modules. While the source code of all modules is typically stored in
standard revision control systems, they may be of varying type, hosted by
various research groups, spread across the world. Cactus provides convenience
tools to automatically assemble a complete source tree from a given list of
modules~\cite{Seidel:2010aa}, and to compile on a large list of known
supercomputers and standard environments~\cite{Thomas:2010aa}.

This component-based design makes it possible to provide fairly
interesting high level features which include, e.g., (1)
externalizing the driver into a component as described above, (2) a
generic time integration module that provides high-order coupling
between independently-developed physics modules, or (3) an integrated web
server for monitoring simulations that offers functionality similar to a
debugger~\cite{Korobkin:2011tg} or Web 2.0 integrations~\cite{Allen2009c}.

\emph{Kranc}~\cite{Husa:2004ip,Kranc:web} is a Mathematica-based tool
that generates complete Cactus components from equations specified in
Mathematica syntax. In particular, Kranc allows tensorial equations to
be written in a compact way employing abstract index notation. Kranc
is able to apply a set of high-level optimizations that compilers are
typically unable to perform, especially for large compute kernels
(loop fission/fusion, SIMD vectorization). Kranc generates both C++
and OpenCL code.



\subsection{Enzo\label{sec:enzo}} 

Enzo's SAMR 
supports arbitrary block sizes and aspect ratios
(although blocks must be rectangular solids, and there are some
practical limitations on their sizes and locations).  Adaptive
time-stepping is used throughout the code, with each level of the grid
hierarchy taking its own timestep.  This adaptive time-stepping is
absolutely critical to the study of gravitationally-driven
astrophysical phenomena, since the local timescale for evolution of
physical systems typically scales as $\rho^{-0.5}$, where $\rho$ is
the local density. 

Enzo uses C++ for the overall code infrastructure and memory management,
and typically uses Fortran-90 for computationally-intensive solvers.
Within the code, the fundamental object is the ``grid patch,'' or
block, and
individual patches are composed of a
number of baryon fields, as well as particles of a variety of types.
The non-local solvers for gravity and implicit flux-limited diffusion
require substantial amounts of communication -- gravity solves
currently use a FFT-based method on the root patch, and a multigrid
relaxation-based method on subpatches.

Threading is supported at block-level and deeper level, 
but the improvement in performance has been marginal with deeper
threading compared to block-level parallelism. 
In addition, several of the hydrodynamic and magnetohydrodynamic
solvers have been ported to graphics processing units (GPUs), with
substantial speedup seen in situations where this physics dominates
the computational time (e.g., driven compressible MHD turbulence).
Using the hybrid-parallelized (MPI+OpenMP threads) version of Enzo,
nearly-perfect weak scaling up to 130K cores on a Cray XK5 has been
observed when the code is used in its unigrid (non-AMR) mode, and
reasonable scaling of up to 32K cores has been observed on Blue Waters
(a Cray XK7 machine) using AMR with adaptive time-stepping.

A noteworthy feature of the Enzo code is its development process,
which has become completely open source and community-driven.   
The Enzo project was originally developed by the Laboratory for
Computational Astrophysics at UIUC and then UC San Diego, but is now developed and
maintained by a distributed team of scientists.  Code
improvements are funded at the individual PI level by a wide variety of
sources, and user contribution to the source code is heavily
encouraged.  To facilitate this process, Enzo uses the distributed
version control system Mercurial~\cite{mercurial:web}, which allows
for extremely simple forking, branching, and merging of codebases.
Coupled with an extensive answer testing framework~\cite{Enzo2014} and
a formalized system of peer review (managed by explicit ``pull
requests'' from user forks to the main Enzo codebase), this enables
the code to remain stable while at the same time directly
incorporating feedback from non-core developers.   User contribution
is further encouraged by the code structure, which enables the
straightforward addition of new physics modules, particularly those
that are purely local in their impact (i.e., plasma chemistry, radiative cooling, star
formation and feedback routines, etc.).

\subsection{FLASH \label{sec:flash}}
FLASH combines two frameworks, an Eulerian discretized mesh and a
Lagrangian framework \citep{Dubey2012} into one integrated
code. Though most of the other codes discussed in the paper support
Lagrangian particles in some form, a  general purpose framework for
Lagrangian data is unique to FLASH. The
backbone of the overall code infrastructure is component-based. 
Instead of relying upon F90 object-oriented features for architecting
the code, FLASH imposes its own object-oriented framework built using
a very limited DSL and a Python-based configuration tool that together
exploit the Unix directory structure for scoping and
inheritance. FLASH's {\em Grid} unit manages the Eulerian mesh and all
the associated data structures. Support exists for explicit
stencil-type solvers using a homegrown uniform grid package, and for
AMR using PARAMESH and Chombo. FLASH's explicit physics solvers are
completely oblivious to the details of the underlying mesh and can switch
between packages during application configuration. The elliptic and parabolic
solvers need closer interaction with the mesh, therefore applications
that require those solvers usually default to using PARAMESH, which
has the most comprehensive solver coverage in FLASH. 

The physics units that rely on their solvers interacting closely with
the mesh are split into two components; the mechanics of the solvers that
need to know the details of the mesh, but are agnostic to the physics,
become sub-units within the Grid unit, while the sections that are physics-specific
exist in their own separate physics units. Within the Grid unit, a
unified API is provided for the various underlying solvers, some of
which are interfaces to libraries such as Hypre. The unified API 
exploits the provision for co-existence of multiple alternative
implementations of a given functionality to facilitate the use of the most
appropriate solver for a specific application.  This feature also
allows re-use of the solvers for other purposes as needed. 

The Lagrangian capabilities of FLASH have evolved into a framework of
their own because they can be used in multiple ways, both for
modeling physical particles directly and for providing mechanisms used
by other code units. The mechanics of Lagrangian data movement are
employed for implementing a {\em laser drive} for simulating
laser-driven shock experiments, in addition to their original use for
active and passive tracer particles. Similarly, an immersed boundary
method for fluid-structure interaction makes use of mapping and data
movement mechanics to couple the structure with the fluid through
Lagrangian markers.   FLASH has demonstrated scaling up to 130K cores
in production runs and a million way parallelism in benchmarking on
the BG/P and BG/Q platforms \cite{Dubey2013c, Daley2013} respectively. 

\subsection{Uintah} 
Uintah consists of a set of parallel software components and libraries
in a framework that can integrate multiple simulation
components, analyze the dependencies and communication patterns
between them, and efficiently execute the resulting multi-physics
simulation. Uintah uses task-graphs in a similar way to Charm++
\cite{CharmPP}, but with its own unique algorithms. For example
Uintah uses a ``data warehouse'' through which all data transfer takes
place, and which ensures that the user's code is independent  of the
communications layer. Uintah's task-graph structure of the computation
makes it possible to improve scalability through adaptive
self-tuning without necessitating changes to the taskgraph
specifications themselves. This task-graph is used to map processes onto processors
and to make sure that the appropriate communications mechanisms are in  
place. 

Uintah has been used increasingly widely since 2008, and a concerted effort to improve
its scalability has been undertaken \cite{MBBWorkshop} by building upon   
the visionary design of Parker \cite{uintah2}.
Particular advances made in Uintah are advanced scalable AMR
\cite{luitjens, Justin2, JLCONCURRENCY} coupled to challenging multiphysics problems
\cite{IPDPS10,TGRID10}, and a load balancing data assimilation and feedback approach 
which outperforms traditional cost models.
Uintah originally utilized the widely-used Berger-Rigoutsos algorithm 
\cite{bergerrigoutsos} to perform regridding.
However, at large core counts this algorithm does not scale \cite{JLCONCURRENCY}, 
requiring a redesigned regridder for such cases.  
The regridder in Uintah defines a set of fixed-sized tiles throughout
the domain. Each tile is then searched, in parallel, for refinement
flags without the need for communication. All tiles that contain
refinement flags become patches.  This regridder is advantageous at
large scales because cores only communicate once at the end of
regridding when the patch sets are combined. This approach immediately
led to a substantial increases in the scalability of AMR by a factor
of 20x \cite{IPDPS10}. 
The load balancer 
makes use of profiling-based cost estimation methods that utilize time series
analysis.  These methods provide highly accurate cost estimations
that automatically adjust to the changing simulation based upon a
novel feedback mechanism \cite{IPDPS10}.

\begin{center}
\begin{figure}[h]
  \includegraphics[width=3.5in,draft=false]{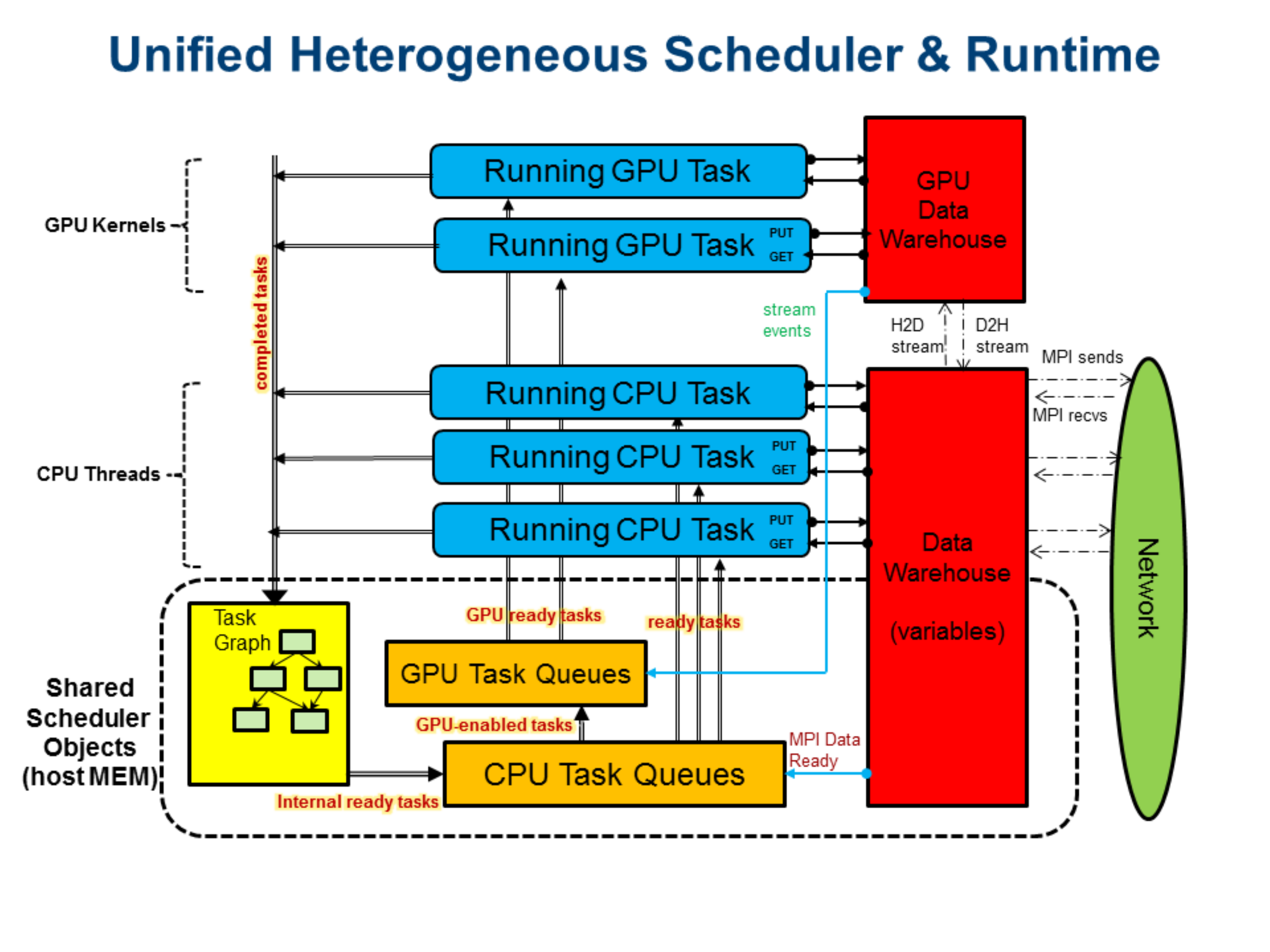}
  \captionof{figure}{Schematic of the  Uintah nodal runtime system \cite{QMWolf}}
  \label{fig:Runtime}
\end{figure}
\end{center}

A key factor in improving performance is the reduction in wait time through
the dynamic and even out-of-order execution of task-graphs
\cite{Qingyu,TGRID10}.
Uintah reduces its memory footprint through the use of a nodal shared memory model
in which there is one MPI rank and one global memory (a Uintah datawarehouse)  per multicore node, 
with a thread-based runtime system used to exploit all the cores
on the node \cite{QMConcurrency}. The task-based runtime system is designed around the premise
of asynchronous task execution and dynamic task management and includes (see Figure \ref{fig:Runtime})the following
features.
Each CPU core and GOU accelerator uses 
decentralized execution \cite{Qingyu} and 
requests its own work from task queques.
A Shared Memory Abstraction through Uintah's data warehouse 
is used to achieve lock-free execution by making use of atomic  operations
supported by modern CPUs so as to allow scheduling
of tasks to not only CPUs but to multiple GPUs per nodes.
Support is provided for multiple accelerators per node and for ensuring that task queues are hosted by the
accelerator\cite{Alan,QMWolf}.
%
The nodal runtime system that makes this possible is shown in Figure~\ref{fig:Runtime}.
Two queues of tasks are used to organize work for CPU cores and accelerators in a dynamic way.
This architecture has now also been used successfully on Intel Xeon Phi accelerators.   
All of these optimizations have resulted in the benchmark Fluid-structure interaction AMR
application's demonstrated scalability to 500K cores and beyond on the Blue Gene
Mira at the DOE's Argonne National Laboratory \cite{QMCSC13}.  

Table \ref{tab:features} summarizes the important framework features
and capabilities in various codes. 

\begin{table}
\begin{center}
\begin{tiny}
\begin{tabular}{|l||l|l|l|l|l|l|} \hline
Feature                   & BoxLib      & Cactus     & Chombo   & Enzo & FLASH & Uintah \\ \hline\hline
Subcycling & optional & optional & optional &required&none& required \\\hline
Timestep & same as  & independent & same as  & independent & & same as  \\
ratio & refinement  & & refinement  &  & & refinement \\\hline
Elliptic &PETSc/& user      &PETSc/ & Hypre &Hypre/ & PETSc/\\
Solver  &Hypre/&supplied &native  & native  & native & Hypre\\
            & Trilinos &          & GMG     & GMG     & GMG &  \\
           & nativeGMG&&&&&\\\hline
GMG & sub or & &sub or & single  & whole &sub or \\
with &  whole & & whole & level & &whole \\
AMR& mesh & & mesh& & mesh &mesh\\\hline 
Spherical AMR & 1D & multipatch & 1D & & 1, 2, or 3D & 1D \\\hline
Cylindrical AMR & 2D & multipatch & 2D & & 1, 2, or 3D & 2D \\\hline
Mixed Dimensions& & &Yes &&&\\\hline
Highest Dimension&& 4D &up to 6D&&&\\ \hline
Block size &variable&variable&variable&variable&fixed&variable\\\hline
Refine factor& 2/4 & 2 & 2/4 &any integer & 2 &any integer \\\hline
parent blocks& not unique & not unique & not unique & unique & unique
& unique \\ \hline
Regridding & tag cells & tag cells & tag cells & tag cells & tag blocks & tag
cells \\
Level & per level & per level & per level & per level & all at once &
all at once \\\hline
Space & Morton &  & Morton &
Piano- & Morton  & Hilbert + fast\\ 
filling curve & &  &  &
Hilbert  &  & sorting \\ \hline
OpenMP & per block & dynamically &per block &per block&
per block& \\
&and loops & tuned loops & & and loops & and loops& \\ \hline
Accelerators&&CUDA&&CUDA&&CUDA\\
&&OpenCL&&&&\\\hline
Parallel I/O & native &HDF5 &HDF5&HDF5&HDF5&HDF5\\
& & &&&PnetCDF&\\\hline
Viz & VisIt/yt & VisIt/yt & VisIt & VisIt/yt &VisIt/yt & VisIt \\ \hline
FSI & & & Embedded & &Immersed & MPM \\
&&&Boundary& &Boundary& Method\\ \hline
Framework&C++/& C/C++&C++&C++&Fortran&C++\\
language&Fortran&&&&&\\\hline
User Module&Fortran&C/C++&Fortran&Fortran&Fortran& \\
language&&Fortran&&&&\\ \hline
\end{tabular}
\caption{\label{tab:features} A summary of features in the SAMR codes and frameworks. In the above
table GMG stands for ``Geometric Multigrid,'' and FSI stands for ``Fluid Structure Interaction.''
The ``Spherical'' and ``Cylindrical'' AMR columns specify whether the AMR structure understands
something other than logically rectangular regions. The entry ``multipatch'' denotes that Cactus
can cover spherical or cylindrical regions by piecing together distorted but logically rectangular regions.
}
\end{tiny}
\end{center}
\end{table}

\section{Performance Challenges \label{sec:challenges}}
Use of SAMR provides an effective compression mechanism for the
solution data by keeping high resolution only where it is most
needed. This data compression comes with certain costs; the
management of mesh is more complex with a lot more meta-data, and
good performance (scaling) is harder to achieve. The design space is
large as observed from the variations found in the SAMR codes.
Some of the performance challenges are inherent in using SAMR,
for example even load distribution among computational resources. Some
others such as memory consumption are artifacts of design choices. 
In this section we discuss the impact of design choices on code 
manageability and performance. 

The dominant design difference between SAMR codes is in the way the
grid hierarchies are managed. Logically the simplest to manage is the
tree with clear parent-child relationship, with more flexibility
proportionately increasing the complexity. The tree structure
combined with the constraint of having identical number of cells in
each blocks makes FLASH's meta-data the easiest to manage and
compress. Not surprisingly, FLASH was among the first SAMR codes to
eliminate redundant replication of the meta-data everywhere. Local
tree views are easy to construct and do not consume too much
memory. The disadvantage is that compression of solution data is less
effective with many more blocks being at higher resolution that they
need to be. All of the other frameworks keep their refinement more
limited and are therefore extremely efficient in solution data
compression. More importantly keeping the extent of compression as a
tunable parameter  gives them a great deal of flexibility in resource
usage. The disadvantage is that devising distributed meta-data, or
even compression of meta-data is more difficult. Enzo constrains
its refinement by insisting on a single parent at the coarser
level for each patch in the finer level, thereby allowing some
assumptions in the meta-data. Other codes such as Chombo and BoxLib 
do not place such constraints in the interest of allowing maximum
flexibility in tagging cells for refinement. In general, more  
flexibility in refinement options translates to more complex
meta-data. Chombo, for example, has a mechanism for compressing the
meta-data, but that imposes restrictions on the size of
individual boxes \cite{van2011petascale}. Replicating meta-data makes
management of the mesh cheaper in communication costs, but more
expensive in memory cost. The scaling limitations of replicating
meta-data are well known, though Chombo team's contention is that
distributed meta-data has been a premature optimization for
patch-based frameworks. This is because MPI rank count and local fast
memory are growing proportionately on all vendor roadmaps. However, 
the straight meta-data replication is unlikely to persist in the
current form in future versions of these codes.  The code that has
gone the farthest in overcoming this limitation is Uintah, which is
also ahead of other codes in embracing newer programming models.      

Achieving balanced load distribution is difficult for all of these
codes not only because of AMR, but also because they support
multiphysics, where different solvers have different demands on the
communication and memory infrastructure.  For example, if specialized
physics such as evaluation of reaction networks is very localized,
some blocks will have much more computational work than
others. Sometimes weighting blocks by the amount of work helps, 
but not if different solvers dictate conflicting weighting on the same
blocks. Similarly, with subcycling finer grids do much
more work, therefore appropriately weighting the blocks and distributing
work becomes harder. One possible solution is to distribute the load on
a per level basis on all the available processors. This achieves load
balance at the possible cost of somewhat longer range communications
for filling the guard cells, which is not a huge cost on moderate
processor count. When there is no subcycling, weighting for load per
block is easier and a global all-levels load distribution gives
reasonable results. Here the disadvantage is that coarser levels do
redundant work because the timestep is usually dictated by the finest
level. For all of these reasons performance and scaling have been
major concerns for SAMR codes and all frameworks adopt some tunability
for performance. For example ``grid-efficiency'' in patch based meshes
allows for trade-off between efficiency of solution data compression
and load balance. Similarly parameters like maximum block size, the
frequency of regridding etc give some tools to the users to optimize
their science output within given resource constraints. A case study of
this kind of optimization can be found in \cite{Dubey2013c}.

\section {Future Directions \label{sec:future}}
As we anticipate the move to architectures with increasingly more
cores and heterogeneous computational resources per node, both
algorithms and software need to evolve. Uintah is perhaps ahead
of all the other codes in exploiting newer approaches in programming
abstractions. The future plans for SAMR and other multiphysics codes
are trending towards removing flexibility from some parts while
adding flexibility to other parts. For example a common theme
among many patch based codes is to move towards a fixed block size to
allow for easier distributed meta-data management. Chombo is
experimenting with this approach, while Enzo is considering a
transition to octree based  code similar to FLASH to avoid memory
fragmentation, difficulty in optimizing solvers, and difficulty in
load balancing.   

Another common theme is more fine-grained specification of tasks
through some form of tiling (in FLASH, Chombo and BoxLib ?) and higher
level specification of computational tasks or workflow items in the
schedule (Cactus and Uintah) to improve both  parallel efficiency, as
well as safety by reducing programming errors. Dynamic task scheduling
already exists in Uintah, others such as Chombo, Cactus and Boxlib are
actively working with researchers in runtime systems to bring it
into their frameworks. The use of domain specific languages is
also under consideration, and is in different degrees of adoption by
different packages. Cactus and Uintah already 
deploy some, and looking at further expansion. 
Chombo uses one at C++/Fortran interface, while FLASH uses one only for configuration
purposes. Code transformation and auto-tuning are under investigation
as well, with some deployment in Cactus and Uintah. 
Cactus is using \emph{Kranc} \cite{Husa:2004ip, Kranc:web} to convert
high-level mathematical expressions and discretized operators into
code (C++, CUDA, or OpenCL), automatically deriving dependency
information. A DSL for Kranc called \emph{EDL} (Equation Description
Language) exists \cite{blazewiczphysics, schnetter:2013q}, but is not
widely used yet.
Uintah proposes to use  
the Wasatch approach proposed by Sutherland
\cite{Notz2012}, in which the programmer writes pieces of code that
calculate various mathematical expressions, explicitly identifying
what data the code requires and produces/calculates. This code is used
to create  both dependency and execution graphs. Uintah will use an embedded DSL 
called Nebo to achieve abstraction of field operations, including
application of discrete operators such as interpolants and gradients
\cite{might:Earl:2012:Nebo} in addition to the directed acyclic graph
expressions that expose the dependency and flow of the calculation.

Several codes are moving to higher order methods as a part of
their strategy for dealing with future architectures. For example, 
most Cactus-based physics components employ high-order finite
difference or finite volume methods for numerical calculations, which
can readily be implemented via Cactus's block-structured grid
functions. This is being extended to support Discontinuous Galerkin
finite element methods. Similarly, almost all applications using
Chombo are now moving to 5th or higher-order accuracy. For example new
Method of Local Correction potential theory elliptical solver based on
an extension of the original MLC solver described in
\cite{mccorquodaleETAL:2007}, but extended to higher order and
exploiting SIMD and manycore parallelism, has been implemented in
Chombo. Transitioning to higher order methods is also a part of
FLASH's future plan. 

BoxLib and Enzo developers are considering 
a fundamental redesign of their time-stepping algorithms, replacing 
traditional block-structured AMR with region-based AMR. BoxLib's
region-based AMR, of which a prototype has been implemented in the
BoxLib framework, replaces the concept of a ``level'' of data with the
concept of a ``region'' of data.   Regions, like levels, are defined
by their spatial refinement; the difference is that while there is
only one level at any spatial resolution, there may be multiple
regions within a domain that have the same spatial resolution but
different time steps.   This enables more  efficient time-stepping, as
different physical areas of the flow may require the same spatial
resolution but not require the same time-step.  In some ways this
resembles a tree code, in that one can track the regions in a tree
structure.  The use of regions also allows more flexible job 
scheduling of different parts of the calculation.   Enzo's plan is 
to take steps to go from global time-steps to semi-local time steps,
and will restrict the possible timesteps in such a way as to make
bookkeeping more straightforward. BoxLib is also replacing
communication-intensive algorithms with communication-avoiding or
communication-hiding algorithms

\section{Summary and Conclusions}
\label{sec:summary}
The application codes and the infrastructure packages described in this
survey provide a snapshot of high level frameworks
utilized by multiphysics simulations when using block structured SAMR 
techniques. The selected set does not claim to be comprehensive; there
are many more AMR-based codes and infrastructure packages that are in
active use by different communities. Rather, it is representative of
the different approaches, capabilities and application areas served by
AMR. The codes described here share many common characteristics. They
have all been in development for several years, and are publicly
available (see Table \ref{table:download} for the download sites). The
codes also all have active user communities -- their users can either
contribute back to the code base for inclusion in distribution, or
develop their own applications within the framework. 

\begin{table}
\begin{center}
\begin{tabular}{|c|c|c|c|}
  \hline
Release & For More Info / & How to Access & Registration  \\
        & Where to Access &               & Required? \\
  \hline
BoxLib & ccse.lbl.gov/BoxLib & git & N \\
       & ccse.lbl.gov/Downloads & &  \\
  \hline
Cactus & cactuscode.org/download,       & svn, git & N \\
       & einsteintoolkit.org/download & & \\
  \hline
Chombo & commons.lbl.gov/display/chombo & svn & Y \\ 
  \hline
Enzo   & enzo-project.org & hg & N \\
  \hline
FLASH  & flash.uchicago.edu/site/flashcode &  web   & Y \\ 
  \hline
Uintah & uintah.utah.edu                   & svn & N \\
  \hline
\end{tabular}
\end{center}
\caption{Where and how to access the different frameworks.}
\label{table:download}
\end{table}

All the releases have core infrastructure support that provides a
layer of abstraction between the physics solver capabilities and the
nitty-gritty of housekeeping details such as domain decomposition,
mesh management, and IO.  They have interfaces to math solver libraries
for providing more capabilities. Additionally, they all provide
varying degrees of customizability through their frameworks.  All of the
codes have been used on multiple generations of high end HPC platforms
and have demonstrated good performance at scale, as described in their
respective sections. 

The codes included in this survey share a common concern with other
similarly extensive codes bases -- namely, how to position themselves
with regard to the future platform architectures. They are large
enough that customizing them to a specific machine, let alone a class
of machine architecture, is not a realistic option. Furthermore, the
developers of the codes have collectively seen the advantages of
optimizations that are achieved through better engineering of their
frameworks over platform-specific optimizations, which tend to
have shorter life-span in usability. The codes are therefore moving
towards restructuring their frameworks through more abstractions and
design simplifications. They are in various stages of interfacing with the
abstractions that have developed since the time that their frameworks
were originally developed. Uintah is perhaps the most advanced in the
deployment of runtime systems; however, other codes are not far
behind since future architectures dictate the need to eliminate the bulk
synchronous model that most codes currently employ. The convergence of
application needs in the face of a more challenging HPC landscape and the
maturation of technologies such as task-graph-based runtime, embedded
DSLs and code transformation is set to transform the landscape of
multiphysics application code frameworks in the next few years. The
codes described in this survey, because they are critical research
tools for many scientific domains,  are likely to be among the first to
attempt and to successfully go through this transformation.

\section{Acknowledgments}

{\bf BoxLib} Much of the BoxLib development over the past 20+ years has been
supported by the Applied Mathematics Program and the SciDAC program under the
U.S. DOE Office of Science at LBNL under contract No.\ DE-AC02-05CH11231. 
Scaling studies of BoxLib have used 
resources of NERSC and OLCF,  which are supported by the Office of Science of the U.S.
DOE under Contract No. DE-AC02-05CH11231, and
DE-AC05-00OR22725 respectively.

{\bf Cactus} Cactus is developed with direct and indirect support from a number of
different sources, including support by the US National Science
Foundation under the grant numbers 0903973, 0903782, 0904015 (CIGR)
and 1212401, 1212426, 1212433, 1212460 (Einstein Toolkit), 0905046
(PetaCactus), 1047956 (Eclipse/PTP), a German Deutsche
Forschungsgemeinschaft grant SFB/Transregio~7 (Gravitational Wave
Astronomy), and a Canada NSERC grant. Computational resources 
are provided by Louisiana
State University (allocation hpc\_cactus), by the Louisiana Optical
Network Initiative (allocations loni\_cactus), by the US National
Science Foundation through XSEDE resources (allocations TG-ASC120003
and TG-SEE100004), the Argonne National Laboratory, NERSC, and Compute Canada.

{\bf Chombo} Chombo was started in 1998 and has been in continuous development
since then at Lawrence Berkeley National Laboratory,  primarily supported by 
the U.S. DOE Office of Science at LBNL under contract No.\ DE-AC02-05CH11231, 
and supported for a short time by the NASA Computation Technologies Project (2002-2004).

{\bf Enzo} The Enzo code has been continuously developed since 1995 by the
NSF, NASA, and DOE, as well as by the National Center for
Supercomputing Applications, the San Diego Supercomputing Center, and
several individual universities.  Please consult \citet{Enzo2014} for
the full list of funding sources -- notably, Enzo has been funded by
the NSF Office of Cyber-Infrastructure through the PRAC program
(grant OCI-0832662).

{\bf FLASH} The FLASH code was in part developed by the U.S. DOE-supported ASC
/ Alliance Center for Astrophysical Thermonuclear Flashes at the
University of Chicago under grant B523820. The continued development
has been supported in part  by the US DOE NNSA ASC through the Argonne
Institute for Computing in Science under field work proposal 57789,
and by NSF Peta-apps grant 5-27429.

{\bf Uintah}
Uintah was originally developed at the University of Utah’s 
Center for the Simulation of Accidental Fires and Explosions (C-SAFE)
funded by the U.S. DOE, under subcontract No. B524196.
Subsequent support was provided by the National Science Foundation 
under subcontracts No. OCI0721659, award No. OCI 0905068 and by DOE INCITE awards
CMB015 and CMB021 and DOE NETL for
funding under NET DE-EE0004449. Applications development of Uintah has
been supported by a broad range of funding from NSF and DOE.

\pagebreak
\bibliographystyle{elsarticle-num-names} 
 \bibliography{bibtex/chombo-strings,bibtex/chombo,bibtex/boxlib,bibtex/einsteintoolkit,bibtex/cactus,bibtex/flash,bibtex/UintahAMR.bib,bibtex/enzo.bib,bibtex/AMR.bib}

\end{document}